\documentclass[referee]{raa}           
\usepackage{graphicx,times}
\usepackage{natbib}
\usepackage{amssymb,amsmath}
\bibpunct{(}{)}{;}{a}{}{,}

\usepackage[a4paper=true,dvipdfm=true,pagebackref=true]{hyperref}
\hypersetup{pdftitle = The title of my PDF, pdfauthor = My name, pdfsubject= The subject, pdfkeywords = keyword1 keyword2 keyword3} 
\hypersetup{colorlinks = true, linkcolor = green, anchorcolor = red, citecolor = blue, filecolor = red, pagecolor = red, urlcolor = red}

\begin{document}
\title{Flux Emergence in the Solar Active Region NOAA 11158: The Evolution of Net Current}

 \author{P.~Vemareddy\inst{1}, P.~Venkatakrishnan\inst{2}, and S. Karthikreddy\inst{3}}

   \institute{ $^1$Indian Institute of Astrophysics, Koramangala, Bangalore-560 034, India; {\it vemareddy@iiap.res.in} \\
	$^2$Udaipur Solar Observatory, Physical Research Laboratory, Badi Road, Dewali, Udaipur- 313 001, India; {\it pvk@prl.res.in} \\
	$^3$Department of Earth and Space Sciences, Indian Institute of Space Science and Technology, Thiruvananthapuram-695 457, India; {\it karthikreddy311@gmail.com}
}

\abstract{We present a detailed investigation on the evolution of observed net vertical current using a time series of vector magnetograms of the active region (AR) NOAA 11158 obtained from Helioseismic Magnetic Imager. We also discuss the relation of net current to the observed eruptive events. The AR evolved from $\beta\gamma$ to $\beta\gamma\delta$ configuration over a period of 6 days. The AR had two sub-regions of activity with opposite chirality: one dominated by sunspot rotation producing a strong CME, the other showing large shear motions producing a strong flare. The net current in each polarity over the CME producing sub-region increased to a maximum and then decreased when the sunspots got separated. The time profile of net current in this sub-region followed the time profile of the rotation rate of the S-polarity sunspot of the same sub-region. The net current in the flaring sub-region showed a sudden increase at the time of the strong flare and remained unchanged till the end of the observation, while the sunspots maintained their close proximity. The systematic evolution of the observed net current is seen to follow the time evolution of total length of strongly sheared polarity inversion lines in both the sub-regions. The observed photospheric net current could be explained as an inevitable product of the emergence of a twisted flux rope, from a higher pressure confinement below the photosphere into the lower pressure environment of the photosphere.
\keywords{Active Regions, Magnetic Fields; Activity, Coronal Mass Ejections, Non-Potentiality, Electric Current}
}
\authorrunning{Vemareddy, Venkatakrishnan and Karthikreddy}      
\titlerunning{Evolution of Net Current in AR 11158}  
\maketitle

\section{Introduction}
\label{Intro}
One of the most sought after goals of space weather research is the prediction of solar eruptions. After several decades of theoretical and observational work, there is a convergence on the essential fact that magnetic stress of solar active regions (ARs), stored in non-potential fields is one of the likely sources of energy for powering these eruptions. Along with the availability of magnetic free energy, a trigger is also needed to initiate the physical processes leading to the eruption. Emergence of magnetic flux seems to be a prime candidate as a flare trigger. In recent times, it is also recognized that emergence of current carrying flux is capable of providing the impetus in the form of a Lorentz force to create the conditions for magnetic reconnection, followed by magnetic eruption, in an otherwise force free coronal environment \citep{ravindra2011}.

With the advent of high resolution vector magnetographs on board satellites like Hinode and Solar Dynamics Observatory (SDO), the measurement of photospheric vector magnetic fields has attained very great sensitivity, owing to freedom from the degradation of images produced by the earth's atmosphere. In particular, the availability of full disk data at high cadence from the Helioseismic Magnetic Imager (HMI) allows us to monitor the evolution of the vector magnetic fields of all active regions present on the earth-facing disk of the sun at any given time. This gives total coverage of the evolution of the magnetic field prior to and during all solar eruptions, thereby greatly aiding the search for those changes in the magnetic field that could herald a solar eruption. One parameter, derived from a vector magnetogram, that provides a meaningful global measure of magnetic non-potentiality is the net current of a sunspot. The net current can be readily obtained by integrating the electric current density over all pixels of the region of interest.  

The generation of electric current in astrophysical plasma has been clearly explained by \citet{parker1979, parker1996} in terms of the distortion of the magnetic field by external forces applied by a field free plasma. These local distortions of the local magnetic field, result in local sources of the electric current. However, in the case of a completely isolated magnetic flux bundle, confined by the external field free plasma, the net current obtained by summing all normal components of current density over any cross-section of the flux bundle must vanish. Because if it doesn't, then there will be a spill-over of magnetic field beyond the flux bundle, caused by the non-vanishing net current as per the Biot-Savart law. According to this, we should expect the net current flowing across the photospheric layer of a sunspot, embedded in the field free photosphere, to be zero. This prediction, was verified in a large number of quiescent sunspots \citep{venkat2009}.

However, departures from this prediction were seen from very early times \citep{leka1996,wheatland2000}. \citet{parker1996} argued that some departures from neutralization should be expected on account of insufficient spatial resolution of the magnetographs. However a clear evolution of the observed net current from zero value to a large non-zero value was seen during the emergence of magnetic flux in NOAA AR 10930 \citep{ravindra2011}. It was also noticed that the net current in AR 10930 was chiefly contributed by large sections of highly sheared polarity inversion lines (PILs), a result which had already been demonstrated by \citet{falconer2001}. Since such large scale and coherent behavior of the PILs will be relatively immune from the effects of spatial resolution, we need to reconcile the rather simple application of Maxwell's equations for confined and isolated flux bundles \citep{parker1996} with the equally robust observations of \citet{ravindra2011}.

On the other hand, theoretical simulations have come up with different scenarios for producing non-neutralized current, e.g. by the in situ shearing motions of an already emerged flux-rope \citep{aulanier2005, torok2003}, or by the emergence of a twisted flux-rope into a pre-existing field \citep{torok2014}. This compels an observational re-examination of the conditions under which Parker's requirement of neutralized currents \citep{parker1996} can break down.

The AR 11158 provided a unique opportunity to study this problem. This AR emerged on 11 February, 2011 at the heliographic location E33S19 with complex motions, evolving from a $\beta\gamma$ to $\beta\gamma\delta$ configuration over a period of 6 days and showing prolific activity during its disk transit till February 21. Because of its highly eruptive nature, many studies have investigated this AR thoroughly, for example: in the context of X2.2 flare and consequent CME \citep{schrijver2011}, magnetic field and energy evolution \citep{xudong2012}, helicity injection by flux motions \citep{vemareddy2012a}, sunspot rotations and non-potentiality \citep{jiang2012, vemareddy2012b}, transient magnetic and velocity field changes \citep{maurya2012}, localized horizontal field and vertical component of force changes \citep{wangs2012, wangs2014}, collapsing fields in association with X2.2 flare \citep{gosain2012}, convective zone signatures of magnetic fields \citep{chintzoglou2013}, and so on. The single most fascinating aspect about this active region is the emergence of flux with opposite chiralities in two distinct sub-regions, affording two case studies of emergence within a single active region.

In this paper, we study the evolution of the net current for the N and S polarity sunspots of the active region as a whole and then examine the evolution in the two different sub-regions which produced several CMEs and flares. We identify plausible locations that account for the systematic evolution of the non-neutralized, net current. We also monitor the evolution of positive and negative currents within each polarity of each sub region. In addition, we also look at some morphological changes that occurred during this evolution and discuss the relation of net current to these morphological changes. Finally, we try to find whether the different phases in the evolution of the net current could be linked to the observed eruptive events.

We organize this paper as follows. Description of the data and computation of net current is given in Section~\ref{data}. Brief description of the results are presented in Section~\ref{Res} with a discussion of the results in Section~\ref{disc}.
\section{Observational Data and Analysis Procedure}
\label{data}
For the proposed study of evolution of net current, we used a time series of 480 vector magnetograms of AR 11158 obtained from \textit{Helioseismic Magnetic Imager} (HMI, \citealt{schou2012}) on board \textit{Solar Dynamic Observatory}, covering the period February 13-16, 2011. HMI observes the full solar disk in the Fe {\sc i} 6173\,\AA~spectral line with a spatial resolution of 0.5 arcsec/pixel. Filtergrams are obtained at six wavelength positions centered at 6173\,\AA~line to compute Stokes parameters I, Q, U, and V. These are then reduced with HMI science data processing pipeline \citep{hoeksema2014} to retrieve the vector magnetic field using Very Fast Inversion of the Stokes Vector algorithm \citep{borrero2011} based on the Milne-Eddington atmospheric model. The inherent $180^{\rm o}$ azimuthal ambiguity is resolved using the minimum energy method \citep{metcalf1995,leka2009}. Finally, the projection effects in the field components are corrected by transforming them to disk center using cylindrical Equal Area projection method \citep{calabretta2002,hoeksema2014}.

{\bf Computation of Line-of-Sight Flux and Vertical Current:}

The net flux of any magnetic polarity in a region of interest of the AR are computed as

\begin{equation}
\Phi =\sum\limits_{i=0}^N B_z \Delta x \Delta y
\end{equation}
 
where $\Delta x$ and $\Delta y$ are dimensions of pixel size. We consider pixels having greater than 50\,G in this computation of net flux of $B_z$ distribution. The uncertainties (provided by HMI after vector field analysis pipeline) of $B_z$ distribution are propagated to estimate the error limit of the computed flux as 

\begin{equation}
\delta \Phi = \sqrt{ \sum\limits_{i=0}^N \left(\delta B_z\right)_i^2 }\Delta x\Delta y
\end{equation}

According to Ampere's law, the current density can be written as $\mathbf{J}=\frac{1}{{{\mu }_{0}}}\nabla \times \mathbf{B}$, where $\mu_0=4\pi \times 10^{-7}$ Henry\,m$^{-1}$ and {\bf J} has units of $A m^{-2}$. With the observations of vector magnetograms, we can only compute the vertical component of the current density:

\begin{equation}
J_z=\frac{1}{\mu_0}\left( \frac{\partial B_y}{\partial x}-\frac{\partial B_x}{\partial y} \right)
\end{equation}

where the partial derivatives are approximated using three-point lagrangian interpolation procedure. From these distributions of current densities, the net current $I=\sum\limits_{i=1}^{N}{{{({{J}_{z}})}_{i}}}\Delta x\Delta y$ can be estimated by summing over N pixels in the region of interest. Since the noise in the transverse magnetic field is 50\,G, we perform all computations of the current density after setting the threshold of horizontal field strength at 150\,G to avoid inconsistent results. Further, the uncertainties in these horizontal field components (provided by HMI pipeline) are also propagated according to Equation (3) to estimate uncertainty in the vertical current density as given by

\begin{equation}
\left(\delta J_z\right)_{i,j} = \frac{1}{\mu_0}\left(\frac{ \sqrt{\left(\delta B_x\right)_{j-1}^2+ \left(\delta B_x\right)_{j+1}^2 + \left(\delta B_y\right)_{i-1}^2
+\left(\delta B_y\right)_{i+1}^2}}{2\Delta x}\right)
\end{equation}

where i, j refer to pixel indices in x and y direction, respectively. An equal spacing grid size in both x and y-directions is assumed in arriving at the above expression. Similar to magnetic flux, the error limit of the net vertical current in a given polarity over a region of interest is estimate as

\begin{equation}
\delta I=\sqrt{\sum\limits_{i=0}^{N}{(\delta {{J}_{z}})_{i}^{2}}}\Delta x\Delta y
\end{equation}
 
Although computationally expensive, the above procedure is employed on every vector magnetogram at every 12 minute interval and plotted net flux and current with error limits as a function of time.

\begin{figure}[!htb]
\centering
\includegraphics[width=.90\textwidth,clip=,bb=12 29 422 323]{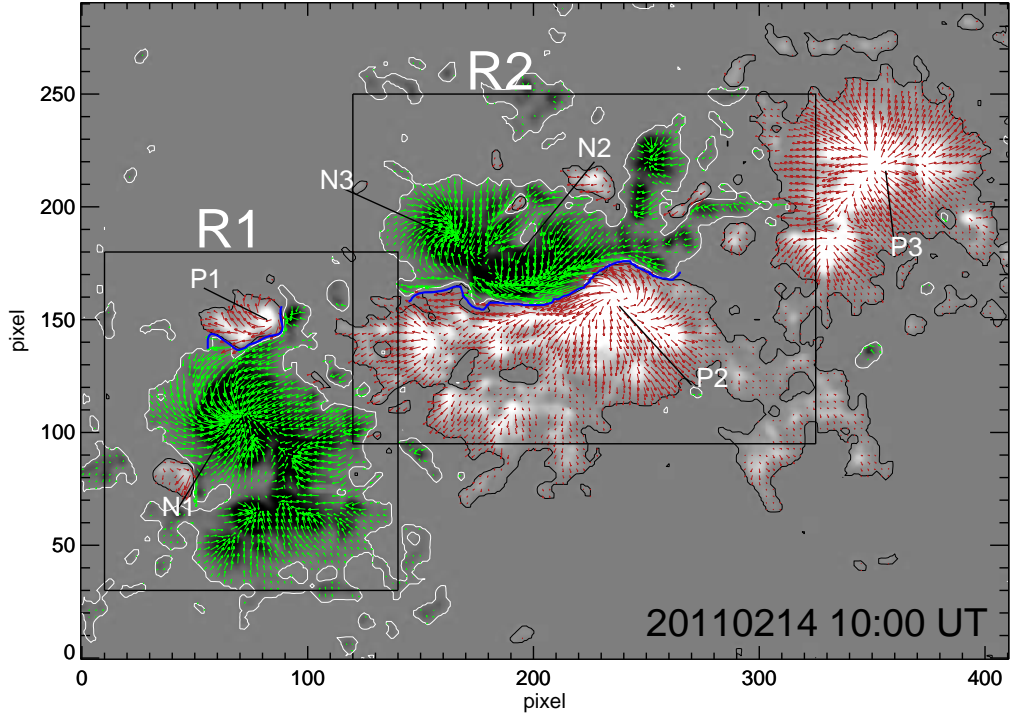}
\caption{Typical Vector magnetogram of AR 11158 at 10:00UT on February 14, 2011. The horizontal field vectors in red (green) are over plotted on vertical component of magnetic field map with iso-contours at 150\,G (-150\,G). The dominant sunspot polarities are marked as P/N* within the rectangular regions of interest R1 and R2 (sub-regions) for further correspondence. The blue solid curves represent the strongly sheared (with shear angle greater than $45^{o}$) polarity inversion lines (PILs) separating major positive and negative vertical flux regions. The field of view is $207\times146$ arcsec$^2$ (1 pixel = 0.5 arcsec).} \label{Fig1}
\end{figure}

\begin{figure}[!htb]
\centering
\includegraphics[width=.99\textwidth,clip=,bb=13 5 584 319]{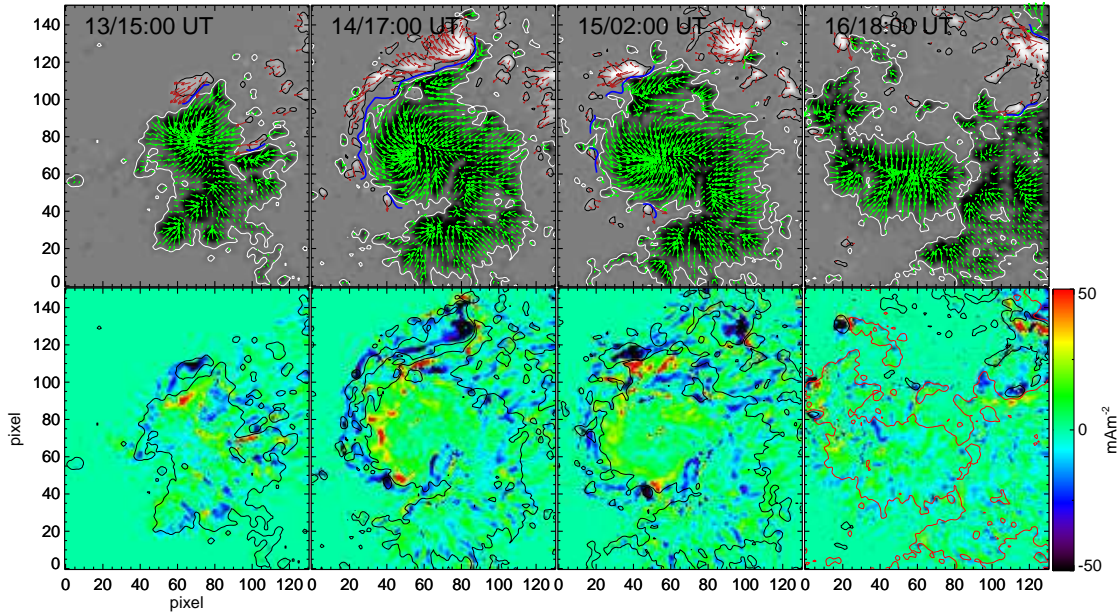}
\caption{\textit{Top row} Evolution of the magnetic field at different epochs of time in sub-region R1. Horizontal magnetic field ($B_h=\sqrt{B_x^2+B_y^2}$) vectors are plotted on the map of vertical magnetic field component. The length of vectors indicates magnitude of $B_h$ and arrow shows direction. The traces of strongly sheared sections of PILs are shown with thick blue curves in each panel. \textit{Bottom row} Distribution of vertical current density ($J_z$) at corresponding times of magnetic field maps. Contours ($\pm150$\,G) of vertical magnetic field are plotted in all panels. All these maps are scaled within $\pm50$\,$mAm^{-2}$ as shown with the color scale.} \label{Fig2}
\end{figure}

\begin{figure}[!htb]
\centering
\includegraphics[width=.99\textwidth,clip=,bb=13 5 595 219]{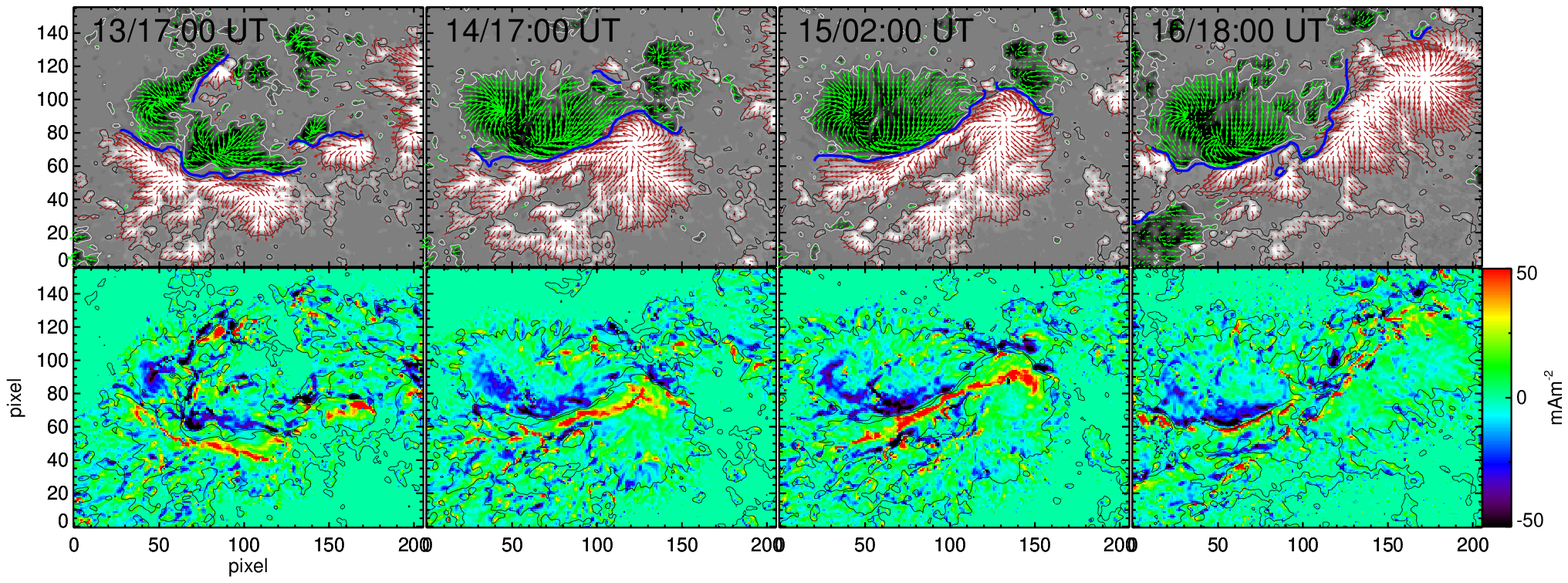}
\caption{Same as Figure~\ref{Fig2} but for sub-region R2.} \label{Fig3}
\end{figure}
\section{Evolution of Magnetic Flux and Vertical Current}
\label{Res}
AR 11158 emerged on February 11, 2011 with prominent, major sunspots appearing on the disc with positive (north) polarities P1, P2, P3 and negative polarities N1, N2, N3 (south) as shown in a typical vector magnetogram in Figure~\ref{Fig1}. The horizontal field vectors are almost parallel to the polarity inversion lines (PILs) between N1, P1 and N2, P2. Such an alignment is known as sheared configuration where field lines become stressed to store magnetic energy. Thus there are two such sub-regions R1 and R2 (shown in rectangular boxes) with high activity. According to the soft X-ray flux information of GOES, this AR produced 15 C, 2 M and one X class flares that are mainly associated with sub-region R2, and many CMEs associated with R1, during 13-16, February, 2011 with continued activity till disk transit on February 21. Because of this high activity AR 11158 became the subject of many studies \citep{schrijver2011,xudong2012,maurya2012,wangs2012,vemareddy2012a,vemareddy2012b,gosain2012}, but excluding any study of net vertical current. In the present study, we will focus on the evolution of magnetic flux and net vertical current in the entire AR and also in the marked sub regions.

The vector magnetograms of R1 and R2 at different epochs of AR evolution are shown in the upper panels of Figures~\ref{Fig2} and~\ref{Fig3}, respectively. These magnetograms show some PILs where the horizontal field vectors are aligned along the PIL. We have calculated the shear angle and identified the pixels that are having strong shear angle (greater than $45^{o}$) in the vicinity of the PIL ($|B_z|<30\,G$). We trace manually such strong shear sections which are shown by blue curves in Figures~\ref{Fig1},~\ref{Fig2}, and~\ref{Fig3}. The length of such sheared sections of the PILs is seen to change at different epochs of the AR evolution.

Along with these magnetograms, the distribution of vertical current density is shown in the lower panel of Figures~\ref{Fig2} and~\ref{Fig3}. The values of vertical current density are spread over a wide range in the entire AR with a typical maximum value of $200\,mAm^{-2}$ in magnitude in both polarities. In particular, intense distribution with large values is located near the interface between P1 and N1 in R1 and between P2 and N2 in R2. Since the positive (negative) polarity P1 (N1) in R1 has a dominant negative (positive) current density distribution, the field associated with these polarities must have dominant negative or left handed chirality. In contrast, positive (negative) polarity in R2 is associated with dominantly positive (negative) distribution of vertical current density, and therefore has positive chirality or right handed sense of twist.

\begin{figure}[!htb]
\centering
\includegraphics[width=0.49\textwidth,clip=,bb=12 5 488 725]{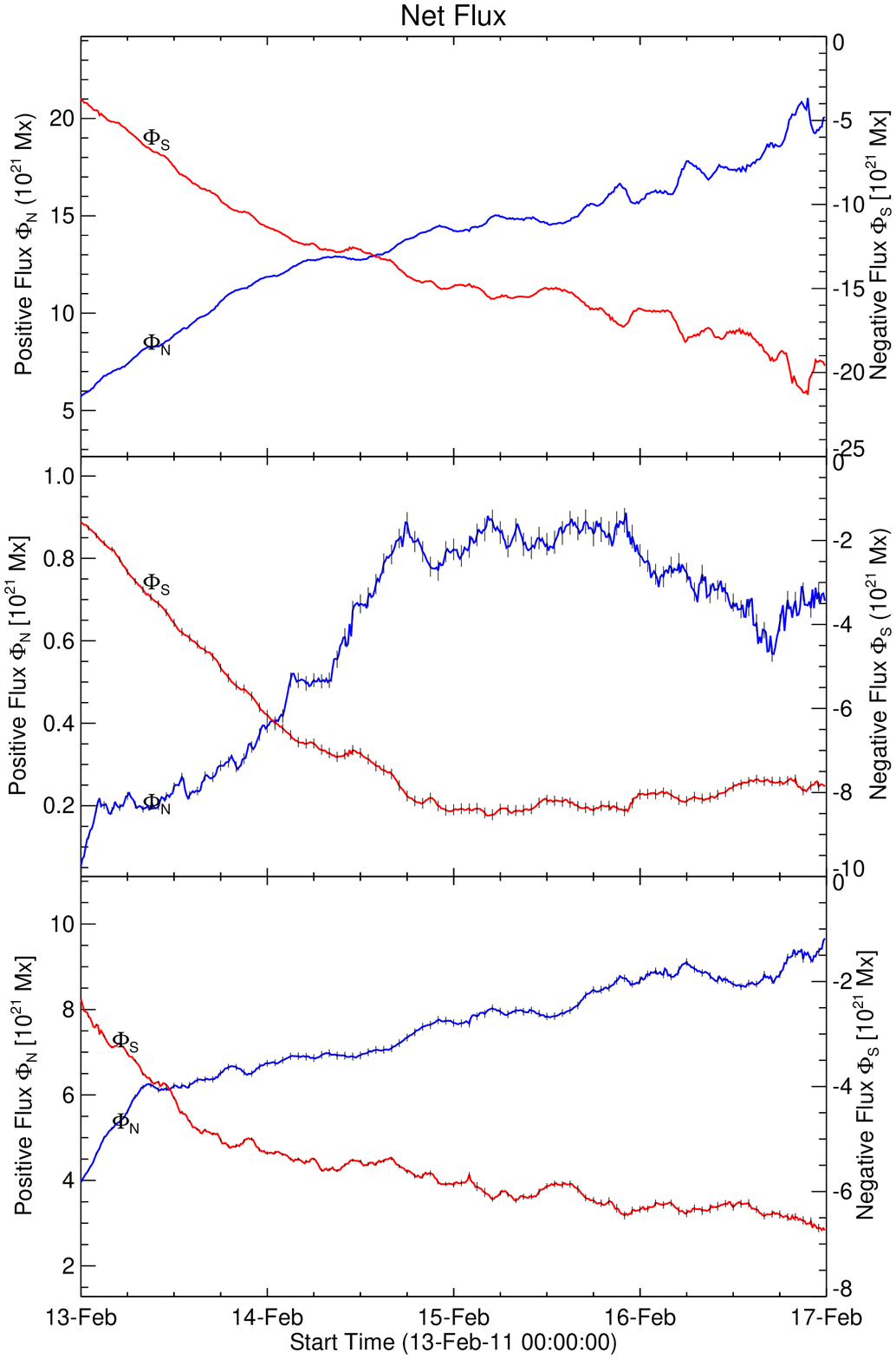}
\includegraphics[width=0.49\textwidth,clip=,bb=22 5 488 725]{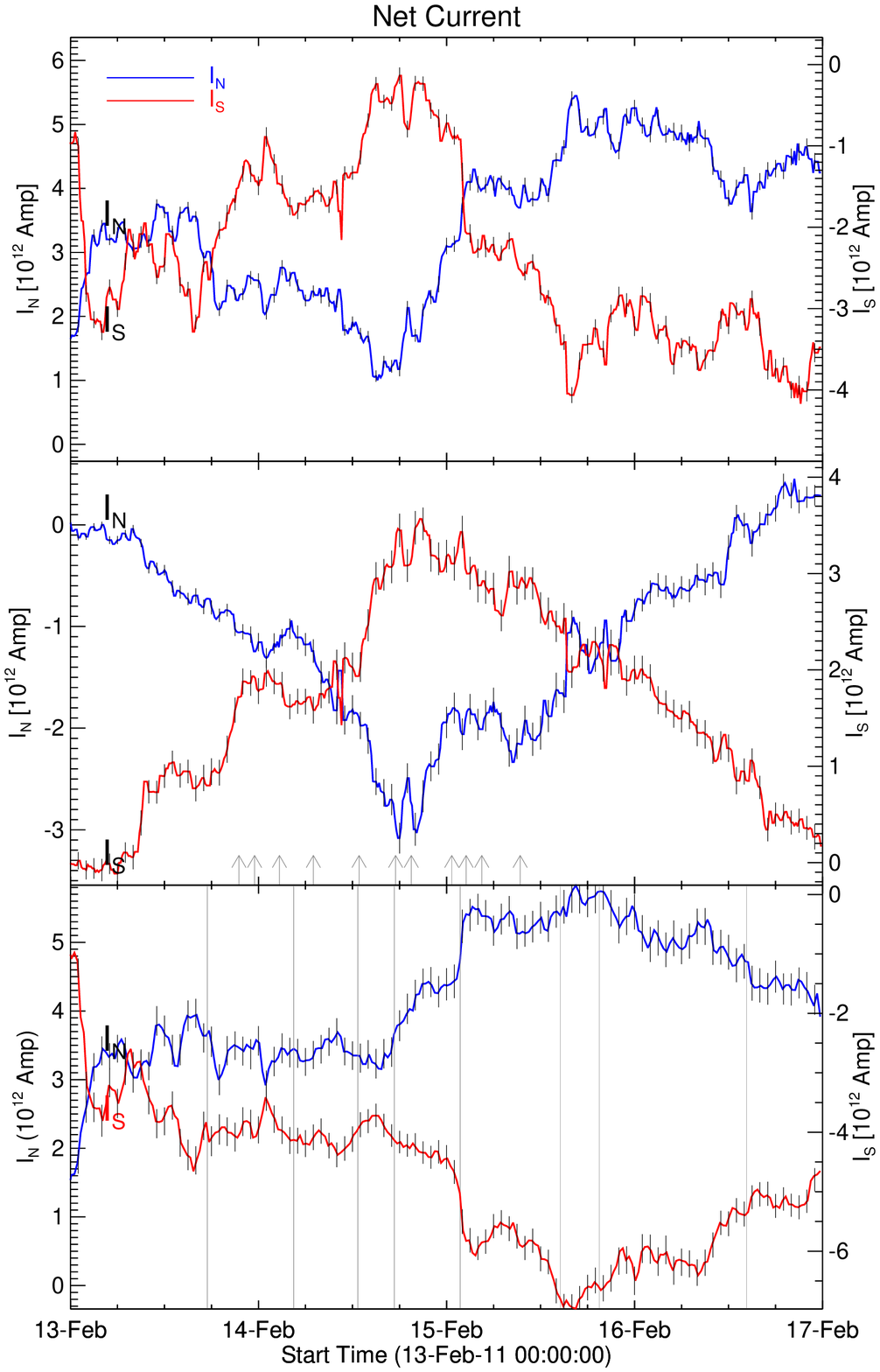}
\caption{Evolution of magnetic flux (vertical current) over entire AR 11158, sub-region R1, sub-region R2 in top, middle and bottom panels of first (second) column, respectively (see Figure~\ref{Fig1}). Vertical lines (Bottom right panel) indicate the initial timings of flares and arrows (middle right panel) are that of CMEs.}\label{Fig4}
\end{figure}

For evolution study, the integrated magnetic flux ($\Phi_N$,$\Phi_S$) and vertical currents ($I_N$, $I_S$) in both polarities are plotted separately, as a function of time, for respectively the entire AR (top panels), region R1 (middle panels), region R2 (bottom panels) in Figure~\ref{Fig4}. Flux in both polarities increases gradually in the AR during four day time period, positive flux at an approximate rate of $1.6\times10^{20}$ Mx/h and negative flux at $1.43\times10^{20}$ Mx/h. This rate is rather high on February 13, at an average $2.5\times10^{20}$ Mx/h in both positive and negative polarities. Most of this flux is contributed by flux emergence in R1 on February 13, where negative flux is emerging at $1.26\times10^{20}$ Mx/h and positive flux at $0.18\times10^{20}$ Mx/h which is about seven times smaller than the negative flux.  As we can notice the uncertainties ranges upto $0.15\times10^{21}$\,Mx without effecting the systematic flux evolution.

The net current over entire AR in each polarity shows sharp increase for 2 hours, then further slow increase till February 13.5, followed by further decrease till February 14.7. From then on, it increases by $5\times10^{12}\,A$ in magnitude till February 15.6 and continues with small variations thereafter. As we can notice the uncertainties ranges upto $0.15\times10^{21}$\,Mx without deforming the systematic flux evolution. This net current profile can be understood in terms of the sum of the profiles of the contributing regions R1 and R2, which evolve differently as discussed below.

In R1, along with the increase in flux, the corresponding currents $I_N$, $I_S$ increased from 13 February, and reached maximum ($I_{N(max)} = 4\times10^{12}A$, $I_{S(max)} = -2.8\times10^{12}\,A$) on February 14.75 and then decreased to minimum value at the end of 16 February. The estimated uncertainties range upto $0.35\times10^{12}$\,A which are small enough to give the right prediction of the net current evolution trend. Note that the sign of current in each polarity is opposite to the sign of flux. The peaking of these currents also coincides with the largest CME originating from R1 associated with the M2.2 flare \citep{vemareddy2012a}.

In region R2, the net current starts with small value, increases rapidly in the first 3 hours of February 13, remains with undulations till February 14.7, and then increases by 75\% ($I_N$ by $3.4-6\times10^{12}\,A$, $I_S$ by $4-7\times10^{12}\,A$) within half a day in both polarities. During this phase of rapid increase in current, there is a sudden increase of net current in both polarities which coincided with a  major CME associated X2.2 flare. Thereafter, the net current decreased by about 30\% till the end of the observations.

The maps of current density $J_z$ for R1 and R2 (Figures~\ref{Fig2} and~\ref{Fig3}) actually show that there are both positive and negative values of $J_z$ within each polarity. When the positive $J_z$ is summed over all pixels with N-polarity magnetic flux, we will get the positive component $I_{N+}$ of the net current and similarly, $I_{N-}$ can be obtained by summing negative $J_z$. Likewise, we can obtain $I_{S+}$ and $I_{S-}$ in  S-polarity flux. We then plot all these quantities with estimated uncertainties as a function of time in Figure~\ref{Fig5}, once again showing 3 rows of panels for complete AR, R1 and R2 respectively. We see immediately that each sub-region exhibits a dominant and non-dominant current, with the dominant current having the same sign as appropriate for the dominant chirality of each sub-region. We also notice that the evolution of the dominant current closely follows the evolution of the net current in each sub-region. The dominant current exhibits positive chirality if summed over both sub-regions and it does not follow the behavior of the net current for the entire region. 

\section{Discussion}
\label{disc}
\subsection{Relation of Observed Net Current with Activity}

The sunspot N1 of the region R1 shows an apparent rotation starting from 14 February. The procedure for the measurement of sunspot rotation with time is described in \citet{vemareddy2012b}. For sake of comparison, we plot the rotation rate ($d\theta/dt$) of the sunspot N1, as measured in \citet{vemareddy2012b}, along with the time variation of $I_S$ in Figure~\ref{Fig6}. The net current in the S-polarity is seen to increase till 20:00\,UT similar to rotational rate profile of N1. We fitted the actual rotation profile ($\theta$) with a theoretical function (Boltzmann-sigmoid) and take the derivative of that fitted profile to derive the rotation rate. Just before the peak rotation of $7^{o}/h$ at 20:00 UT, we observed M2.2 flare with a strong CME at 18:00 UT.  Before this event also, many small CMEs (cf. Table 1 of \citealt{vemareddy2012b}) in the form of mass expulsions we observed associated with this region. It is clear that after 14 February 20:00 UT, the sunspot rotation slows down, accompanied by decrease of the net current.

The observed rotation of the sunspot could be either an actual rotation driven by sub-photospheric twisting motion of the magnetic field or by the emergence of the flux-rope with a gradient of twist along the axis of the flux rope. In the case of the latter scenario, one would imagine that the time rate of apparent rotation would be proportional to the velocity of emergence, for a given axial gradient of the twist. But the rotation rate reaches a peak (cf. Figure~\ref{Fig6}) just when the emergence of flux stops (cf. middle panel of left column in Figure~\ref{Fig4}). This will lead to the rather absurd conclusion that the rise velocity reaches a maximum value when emergence stops. Hence, we reject the hypothesis that the observed rotation is due to the apparent rotation of a twisted flux rope. Rather, we believe that the observed rotation is real. In which case, the increase and decrease in current, accompanying the increase and decrease of the sunspot rotation rate might well be due to the increase and decrease of shear at the PIL caused by the rotation (see also \citealt{suj2008}). Further, since magnetic torque is proportional to the time derivative of the rotation rate of the sunspot (if the moment of inertia is not changed) , we reach an interesting conclusion that the magnetic torque ceases at the time of cessation of flux emergence and thereafter reverses its direction, perhaps indicating a relaxation process.  It may be of interest to note here that many simulations of emergence of a twisted flux tube also show rotation of the ``sunspots'' created by the emergence (cf. \citealt{leake2013})

In addition to the apparent sunspot rotation of N1, there is also the apparent separating motion of P1 away from N1. All these factors probably add to the increase of shear and in turn the current density near the PIL. This increase in current density near the PIL and the consequent increase in net current could well have led to the increase in activity. There are lots of studies supporting the flux emergence as a trigger mechanism of CMEs as is observed in this AR also. Helicity injection calculated by \citet{vemareddy2012a} by tracking the flux motions from this sub-region also followed a similar trend with sunspot rotation profile, with its continuous accumulation into corona leading to CMEs \citep{zhangm2005}.

In the sub-region R2, the origin of currents is different from R1. Although, sunspot P2 also rotated to some extent, its apparent shear motion is more dominant after February 14. Along with the continuous apparent shear motion of P2, the net current also showed increasing trend from 12:00 UT February 14 onwards, with onset of a major CME-associated X2.2 flare at 01:44\,UT on 15 February. An important point to note is that there is only a steady emergence of flux in this region during this time, in contrast to the more rapid increase of net current (See Figure~\ref{Fig4}). As a twisted bundle of magnetic flux-ropes emerges from a higher pressure environment in the sub-photosphere to a lower pressure environment in the photosphere, the magnetic configuration can sometimes dramatically relax, due to a topological change caused by reconnection in the corona. These drastic changes in the magnetic topology could eventually result in a major flare. This scenario of rapid relaxation is consistent with the post-flare increase in the horizontal magnetic field reported in \citet{wangs2012}. The rapid localized increase of horizontal component of the magnetic field could well have resulted in the observed rapid increase in the vertical current.

\begin{figure}[!htb]
\centering
\includegraphics[width=0.99\textwidth,clip=, bb=30 6 597 504]{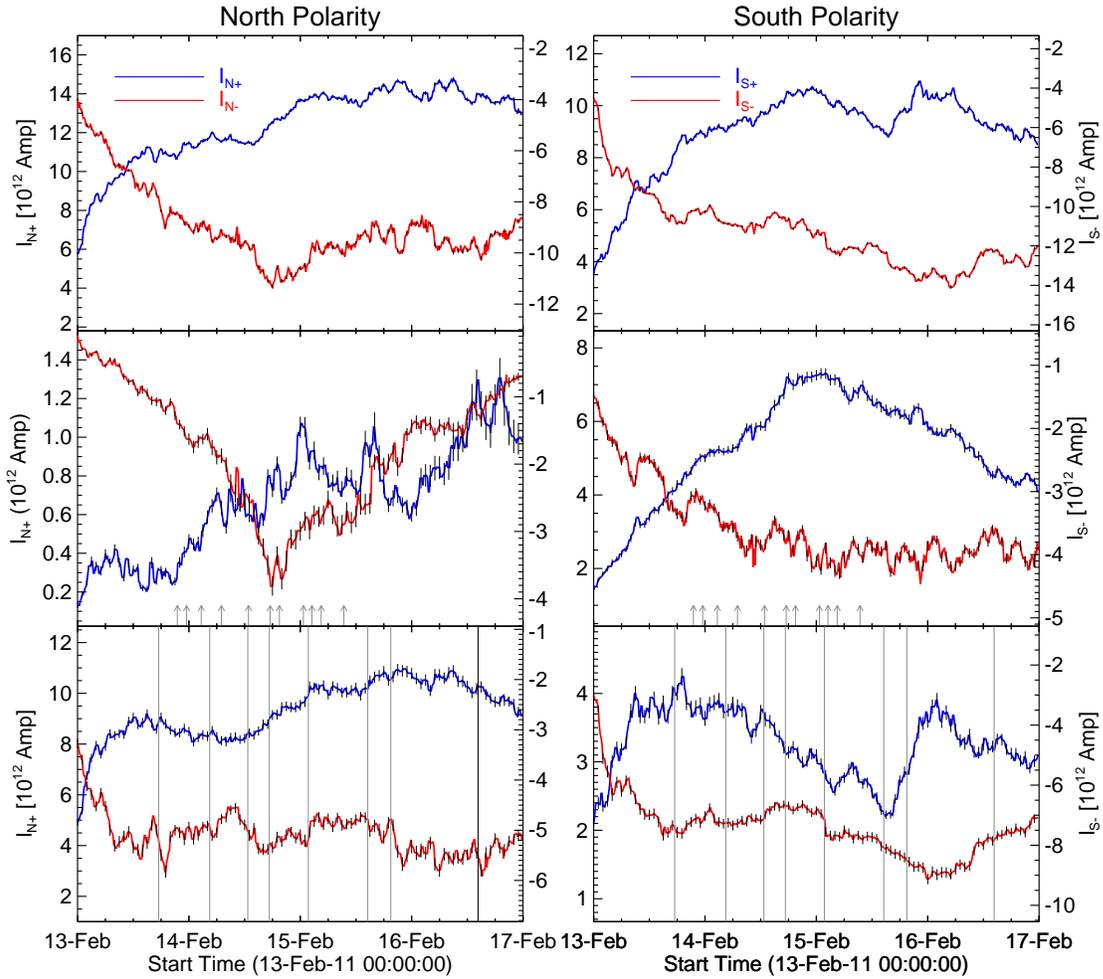}
\caption{Time evolution of individual sign vertical currents in northern (southern) polarity integrated over entire AR 11158, sub-region R1, sub-region R2 in top, middle and bottom panels of first (second) column, respectively.}\label{Fig5}
\end{figure}

\subsection{Systematic Evolution of Observed Net Current}
As mentioned in the introduction, we must expect to find zero net current over a single sunspot on account of the confinement of the flux-rope by external, field-free plasma \citep{parker1996}. \citet{venkat2009} had indeed found this to be the case for a large number of sunspots. A common characteristic of all these sunspots was that they had well defined boundaries with a clear separation of the two polarities having almost negligible horizontal field in the azimuthal direction circumscribing the individual sunspot. However, in the case of AR 10930 \citep{ravindra2011}, the net current was found to increase to rather high values and then there was a decrease. For the AR 11158 studied in the present paper, the net current showed a similar behavior for region R1 while the net current continued to remain large till the end of the observations in R2.

\begin{figure}[!htb]
\centering
\includegraphics[width=0.8\textwidth,clip=,bb=22 7 339 238]{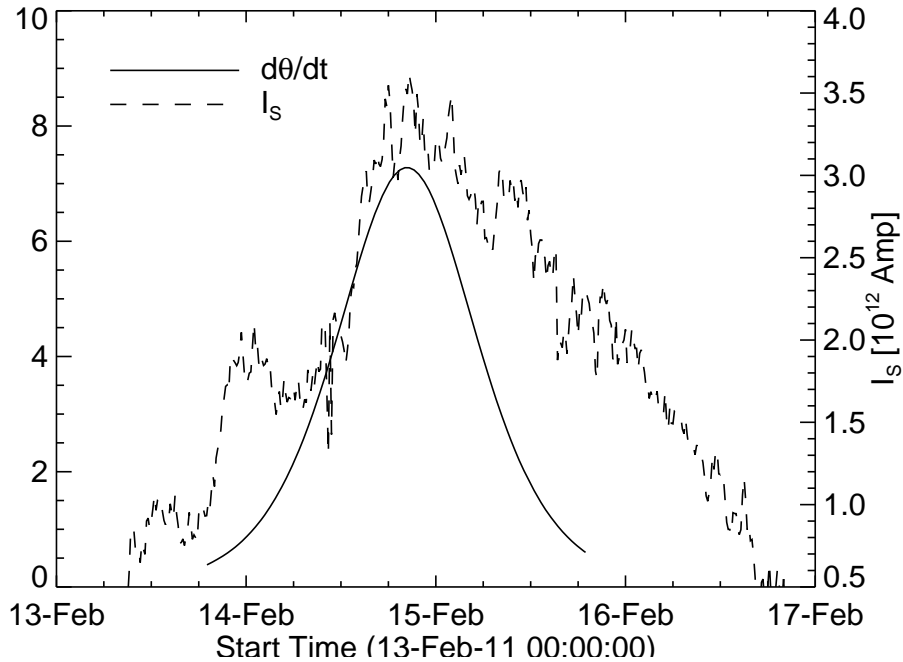}
\caption{Rotation rate of the sunspot N1 plotted with respect to time. For a comparison, the net vertical current from the negative flux in R1 ($I_S$) is also plotted with y-axis scale on the right side.}\label{Fig6}
\end{figure}

\begin{figure}[!htb]
\centering
\includegraphics[width=0.8\textwidth,clip=,bb=17 7 360 417]{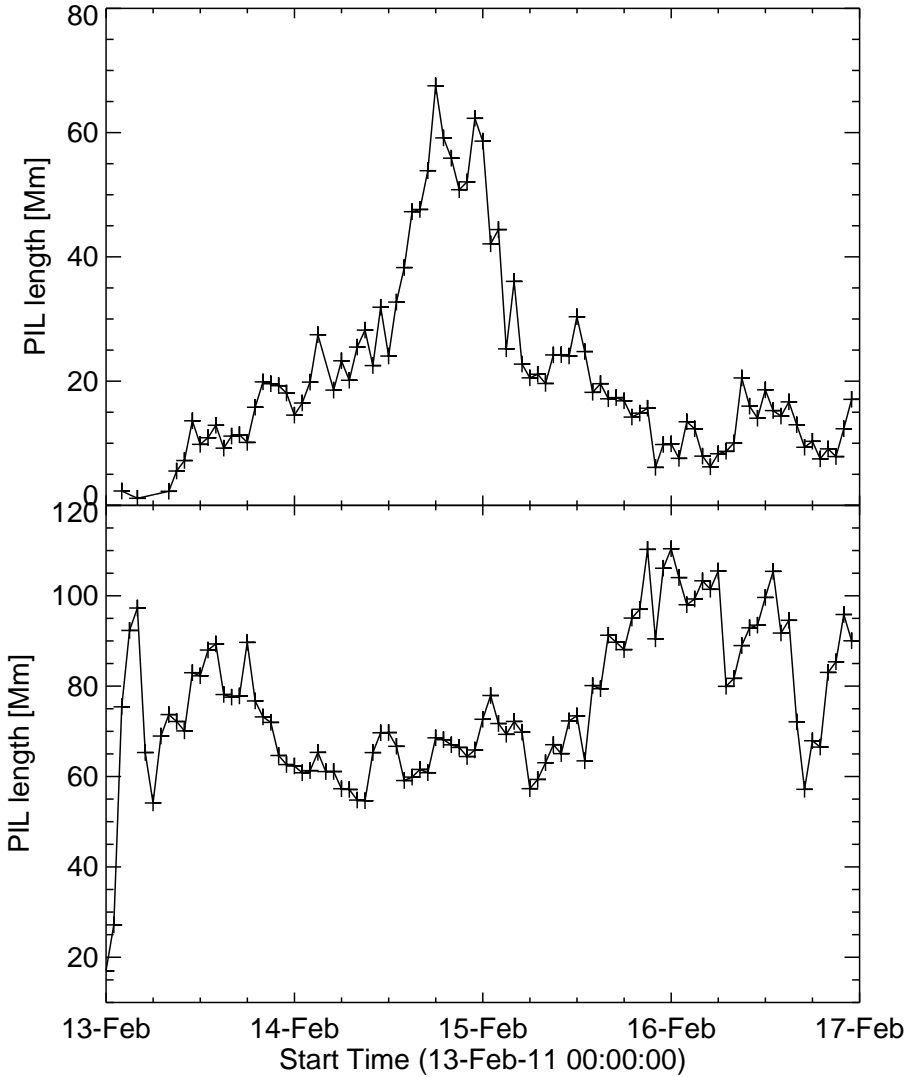}
\caption{Total length of strongly shear PILs in sub-region R1 (top panel) and sub-region R2 (bottom panel) plotted with respect to time.}\label{Fig7}
\end{figure}

Now the question remains as to why the observed non-neutralized current should evolve in such a systematic manner. We can obtain a clue by studying the sheared portions of the PILs of AR 11158. The strong ($>45^{o}$) sheared portions of PIL were traced manually and the lengths of such strongly sheared segments (SSS) were calculated in the sub-regions R1 and R2. We have shown the time variation of total length of SSS from sub-region R1 (top panel) and R2 (bottom panel) in Figure~\ref{Fig7}.  Although manually followed, the expected error can goes upto 6-7 pixels (~2.5Mm) while connecting the strong sheared pixels along the PIL at a given time. In the case of R1, we find that the length of SSS initially increases during 13 to 14 February 2011, then decreases from 14 to 16 February 2011. This trend matches very well with the evolution of net current in R1 (Figure~\ref{Fig4} middle panel). It is interesting to note in this context, that the rotation rate (Figure~\ref{Fig6}) also follows more or less the time variation of the total length of SSS. This reinforces the possible role played by sunspot rotation for an increase in the length of SSS. Likewise, in the case of R2 (Figure~\ref{Fig3}), the changes in the length of SSS (bottom panel of Figure~\ref{Fig7}), matches very well with the evolution of net current in R2 (bottom panel of Figure~\ref{Fig4}). In this case, it is the shearing motions that could have led to the increase in length of SSS. In fact, the correlation of the net current with the length of SSS was already seen in MSFC vector magnetograms \citep{falconer2001}. Physically, this correlation can be explained by the fact that the strongly sheared portions of the PIL contribute maximally to the net current through $\int{\bf B}.d{\bf l}$.

Now, let us address the issue of why Parker's expectation of neutralized currents is not borne out by the observations of non-neutralized net current in emerging flux regions. To get some insight into the physics of the problem, let us look at some recent simulations of \citet{torok2014}, which show that although the confinement of a twisted flux rope by plasma could well neutralize the net current below the photosphere, the situation dramatically changes after the emergence. These simulations clearly show the onset of non-neutralized net current at the start of emergence, which reaches a maximum value when flux emergence stops; thereafter net current decreases asymptotically to a lower value of net current. These simulations also show that the partially emerged flux tube expands laterally such that there is hardly any field free plasma in between the two ``spots''. In such a case, there is no way that one can draw a contour around any one ``spot'' which can remain completely in field free plasma, such that the contour  $\int{\bf B}.d{\bf l}$ can vanish and show zero net current. Even in the absence of a field free interface, if the horizontal field vector lies perpendicular to the PIL, even then we would have zero contribution from the contour integral to give zero net current. But the simulations show the development of strong shear at the PIL which would definitely contribute a net current. Also, the strength of this current is seen to increase with increase in the length of sheared portion of the PIL, a result which closely matches our own observational evolution of net current in step with the observed length of the strongly sheared section of the PIL. It is another question as to why the simulations always showed an increase in shear as a consequence of the emergence. It most probably has to do with the fact that the initial condition has a twisted flux tube confined by high pressure plasma. When the tube emerges, the expansion into a lower pressure environment probably has a rotational component that produces the sheared field at the interface. \citet{leake2013} indeed talk about the transfer of twist from the convection zone into the photosphere, while \citet{longcope2000} discuss the propagation of a torsional alfven wave from the interior to the surface.

As remarked in \citet{torok2014}, it is yet to be seen whether the increase in shear at the PIL is a consequence of the change in connectivity, or whether the shearing motions/rotation of the sunspots leads to the increase in the length of the sheared section of the PIL. Whatever the cause and effect relation between increase in PIL shear and reconnection may be, the final result is the production of net current. One more interesting similarity between the simulation and our observation is the effect of flux emergence on the evolution of the dominant current in R1 and R2 (Figure~\ref{Fig5}), which is consistent with the evolution of the direct current, $I_d$ in the simulations of \citet{torok2014}.


In conclusion, we find that Parker's expectation of a neutralized current in an individual sunspot, is valid only for the evolution of a twisted flux bundle with a field free interface between the two spots. The situation changes dramatically, when the flux of one sunspot emerges into an environment with lower confining pressure, close to another sunspot with opposite polarity. The consequent expansion of the twisted flux tube into the domain of a neighboring sunspot will produce a significant length of strongly sheared PIL without any field free plasma in between. It is also possible that this impact of the two legs of a twisted flux tube can drive reconnection and lead to changes in field connectivity. In which case, appearance of net current in the observed field indicates a possible change in the field connectivity in a slow evolution of magnetic fields in the active region. The changes in field connectivity would, in parallel, lead to increased probability for solar eruptions. 


\begin{acknowledgements}
The authors thank the anonymous referee for the comments which improved the manuscript. The data have been used here courtesy of NASA/SDO and HMI science team. We thank HMI science team for making available the processed vector magnetogram data to the science community. We thank Tibor Torok and Piyali Chatterjee for useful clarifications about simulations of flux emergence. P.V.R is supported by INSPIRE grant under AORC scheme of Department of Science and Technology. 
\end{acknowledgements}
  
\bibliographystyle{raa}

\end{document}